\documentclass[%
 reprint,
 showkeys,
superscriptaddress,
 amsmath,amssymb,
 aps,
]{revtex4-2}

\usepackage{graphicx}
\usepackage{siunitx}
\usepackage{hyperref}
\usepackage[version=4]{mhchem}

\def\pitaffil{%
	\affiliation{%
		Physikalisches Institut, Center for Quantum Science (CQ) and LISA$^+$,
		Universit\"at T\"ubingen, Auf der Morgenstelle 14, D-72076 T\"ubingen, Germany
		}
	}
\def\baselaffil{%
	\affiliation{%
		Department of Physics, University of Basel, 
		4056 Basel, Switzerland
	}
}

\def\sniaffil{%
	\affiliation{%
		Swiss Nanoscience Institute, University of Basel, 
		4056 Basel, Switzerland
	}
}
\makeatletter
\newcommand{\CuOSeO}{\ce{Cu2OSeO3}\ }
\newcommand{\CuOSeOdot}{\ce{Cu2OSeO3}.\ }

\newcommand{\autorefsub}[2]{%
  \hyperref[#1]{\autoref*{#1}(#2)}%
}
\makeatother

\AtBeginDocument{\def\equationautorefname~#1\null{Eq.~(#1)\null}}
\AtBeginDocument{\renewcommand\figureautorefname{Fig.}}
\makeatletter
\newcommand{\Autoref}[1]{%
	\begingroup
	\renewcommand\figureautorefname{Figure}%
	\autoref{#1}%
	\endgroup
}%
\makeatother

\begin{document}

\preprint{APS/123-QED}

\title{Spatial resolution and point spread function of high-resolution scanning SQUID microscopy probes}

\author{Jan Ullmann}
\pitaffil
\author{Katharina Kress}
\baselaffil
\author{Timur Weber}
\pitaffil
\author{Boris Gross}
\baselaffil
\author{Daniel Jetter}
\baselaffil
\author{Reinhold Kleiner}
\pitaffil
\author{Martino Poggio}
\baselaffil \sniaffil
\author{Dieter Koelle}
\pitaffil

\date{\today}

\begin{abstract}
Superconducting quantum interference devices (SQUIDs) show exceptional sensitivity to magnetic flux. In scanning SQUID microscopy (SSM), SQUID size and its distance to the sample are minimized in order to map weak magnetic fields with best possible spatial resolution. SQUID-on-lever (SOL) architectures have proven especially effective as SSM probes due to their small sensor size, robustness, and ease of integration with conventional atomic force microscopy hardware. In order to optimize magnetic microscopy carried out with SOL probes and to accurately reconstruct the magnetic fields that they measure, it is essential to know their point spread function (PSF). The size and shape of this PSF are determined by magnetic flux focusing effects, which depend on the characteristic length-scales of the superconductor and the sensor geometry. By simulating the coupling to sources of magnetic flux, this work provides a mathematical description of the SOL PSF, which contains a full description of the probe's magnetic sensitivity and spatial resolution. We then use measurements of a single magnetic skyrmion to measure the magnetic flux response of a real SOL. We demonstrate excellent agreement with flux responses that are obtained from simulations, thereby confirming the calculated PSF and spatial resolution.
\end{abstract}

\maketitle


\section{Introduction}

Scanning superconducting quantum interference device (SQUID) microscopy (SSM) has provided an effective means of mapping magnetic field patterns for more than three decades~\cite{Kirtley1995, Kirtley1999, Christensen2024}. 
Still, with growing interest in nanometer-scale imaging of magnetic features, including nanostructures~\cite{Vasyukov2018, Wyss2019}, complex oxide interfaces~\cite{Bert2011,Christensen2019}, topological insulators~\cite{Lachman2015,Nowack2013,Ferguson2023}, chiral magnets~\cite{Marchiori2024}, two-dimensional materials~\cite{Noah2023,Zur2023,Vervelaki2024,Bagani2024,Tschirhart2021,Uri2020NatPhys,Uri2020Nat} and various types of superconductors~\cite{Persky2022AnnuRev,Kirtley2010,Jelic2017,Kremen2018,Persky2022Nat}, there is a strong demand for improving the spatial resolution of SSM.

Today, two prominent approaches enable high-sensitivity SSM with high spatial resolution.
The first is the pickup loop susceptometer, or SQUID-on-chip, which is based on a lithographically defined superconducting circuit~\cite{Kirtley2016RSI}.
These devices include a micrometer-scale pickup loop at the corner of the chip, on-chip circuitry for flux modulation and feedback, and a coil concentric to the loop for susceptibility measurements. 
They are robust and can be produced on the wafer scale. 
However, due to limitations in the pickup loop size, spatial resolution is typically in the micrometer range and maximum operating fields top out around \qty{10}{\milli\tesla}, because of the \ce{Nb}/\ce{AlO_x}/\ce{Nb} Josephson junctions (JJs)~\cite{Huber2008}. 
The second approach is the SQUID-on-tip, in which a SQUID loop is fabricated at the apex of an elongated quartz pipette via a three-step directional deposition of superconducting material~\cite{Finkler2010,Vasyukov2013}. 
These probes can have spatial resolutions better than \qty{100}{\nano\meter} and can operate at fields beyond \qty{1}{\tesla}, because of the small diameter of their SQUID loop and the Dayem bridge JJs which form them, respectively. 
By virtue of the proximity of the JJs to the sample of interest, they can even be used for sensitive thermal microscopy~\cite{Halbertal2016}. 
However, SQUID-on-tip probes are fragile, and the fabrication process precludes scaling or the inclusion of circuitry such as a modulation coil for controlling flux or measuring susceptibility.

A new approach to reach comparable or even increased spatial resolution is the SQUID-on-lever (SOL)~\cite{Wyss2022}.
Here, a cantilever designed for atomic force microscopy (AFM) is first coated with a superconducting material; a nanometer-scale SQUID is then patterned at the tip of the lever via focused ion beam (FIB) milling. Using He-FIB, effective SQUID diameters well below \qty{100}{\nano\meter} can be achieved, resulting in magnetic probes with spatial resolutions of the same order~\cite{Weber2025}. 
The significant advantages of this approach have recently inspired other cantilever-based SQUID architectures, including tapping-mode SQUID-on-tips utilizing proximity junctions~\cite{rog_tapping-mode_2026} and wireframe SQUIDs fabricated via corner lithography~\cite{roskamp_nanoscale_2026}.
SOL probes combine the robustness and ease of use of AFM with the favorable properties of both SQUID-on-chip and SQUID-on-tip, i.e.\ high sensitivity, scalable fabrication, integration of on-tip circuitry such as a modulation line or a third JJ, sub-\num{100}-\unit{\nano\meter} spatial resolution, high-field operation, and the ability to map thermal contrast. 

\begin{figure*}[t!] 
    \centering
    \includegraphics[width=\textwidth]{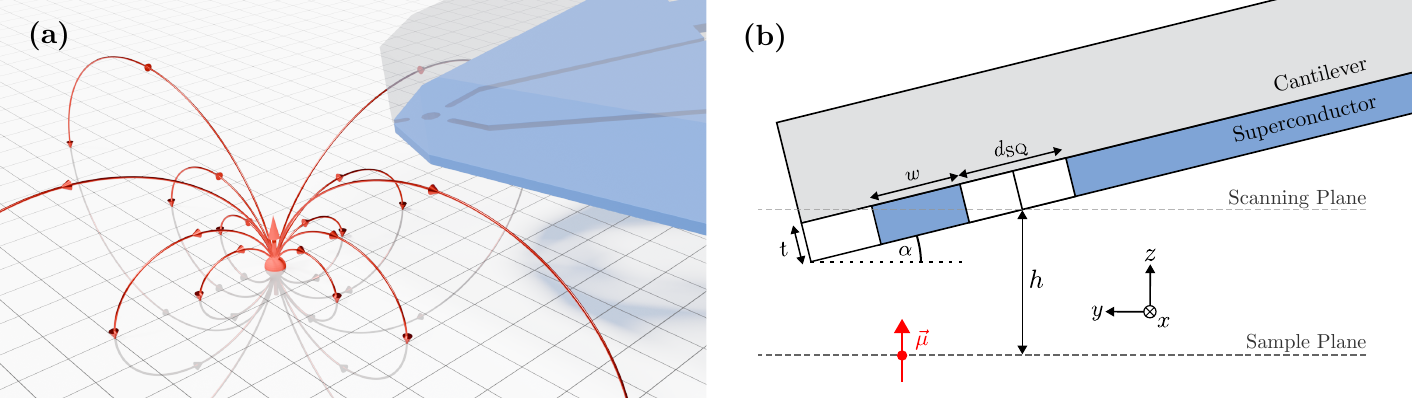}
    \caption{Schematic of SOL. (a) 3D view of the SOL scanning over an idealized point-like magnetic dipole with magnetic moment $\vec{\mu}$. The Nb layer (light blue) on which the SOL circuitry is patterned is on the bottom side of the Si cantilever (semi-transparent). (b) Cross-section of the SOL through its center, with SQUID hole diameter $d_\mathrm{SQ}$, junction width $w$, film thickness $t$ and scanning angle $\alpha$. The operating height $h$ is measured from the center of the SQUID hole.}
    \label{fig:01}       
\end{figure*} 

Quantifying the spatial resolution of SSM devices is challenging and usually reported as an estimate inferred from measurement results~\cite{Pan2021,Dolabdjian2002}. 
Although one can resort to measures such as Rayleigh's or Sparrows's criterion~\cite{Sparrow1916}, full information about the ability to distinguish small features is only contained within the probe's point spread function (PSF). 
The PSF describes how the probe convolves the source image into the measured contrast by including the exact spatial dependence of the probe's response to a point-like source. Fourier analysis of the PSF yields how the probe's sensitivity depends on spatial wavelength, i.e.\ relating the strength of its response to feature size. 
Once the PSF of a probe is known, the measured response can be deconvolved into an image of the source, e.g.\ for SSM, a map of measured magnetic flux can be deconvolved into a map of magnetic field. 

Although measuring a SQUID's PSF is not always feasible, for tip-sample separations larger than the SQUID and the magnetic features, the PSF can be approximated from simple analytical models of the magnetic flux response~\cite{Kirtley2009}. 
As characteristic length scales approach the SQUID diameter or the London penetration depth, simulations considering the properties of the superconducting material as well as the detailed probe geometry are required~\cite{Bouchiat2009, Nagel2011, Linek2024, Kirtley2016SuST}.

Here we simulate coupling factors to model the PSF of SOL probes with the highest spatial resolution to date~\cite{Weber2025}.
We compare our findings with measurements, showing the applicability of the simulation technique. The results quantify the ability of the SOL probes to resolve small spatial features, aid in the deconvolution of images made with these probes, and indicate strategies for further improving spatial resolution.

\section{Simulations}
\label{sec:simulations}

\subsection{SOL geometry}

\Autoref{fig:01} gives an overview of the SOL geometry realized by Weber et al.: it consists of a flat AFM cantilever, whose bottom side is coated with superconducting Nb and patterned by \ce{Ne}- and \ce{He}-FIB milling~\cite{Weber2025}. 
The SQUID hole is centered at the apex of the tip and includes three nearby cuts, forming three Dayem bridge JJs. One cut extends from the front end of the tip towards the SQUID hole. The other two cuts extend down the cantilever body to define three superconducting leads.
Simulations were performed considering a London penetration depth $\lambda_\mathrm{L}$ = \qty{100}{\nano\meter} for Nb with film thickness $t = \qty{50}{\nano\meter}$. During operation, as shown in \autorefsub{fig:01}{b}, the cantilever is tilted by an angle $\alpha = \qty{10}{\degree}$ with respect to the plane over which it is scanned.
At each position in the scan, the SQUID then measures magnetic flux originating from the sample. 
The sample and scanning planes are separated by the SOL operating height $h$.

\begin{figure*} 
    \centering
    \includegraphics[width=\textwidth]{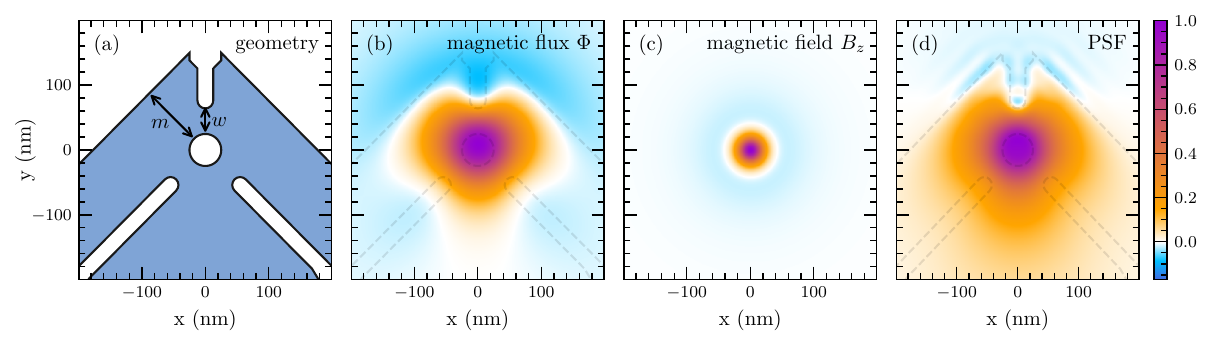}
    \caption{Calculating the PSF of an SOL probe from a magnetic dipole moment. 
    (a) Projection of SOL design with three JJs and diameter $d_\mathrm{SQ}=\qty{50}{\nm}$ onto the sample plane. The junction width is $w=\qty{50}{\nm}$ and the shortest distance between the SQUID hole and the cantilever edge is $m=\qty{100}{\nm}$. The thickness of the Nb film is $t=\qty{50}{\nm}$. 
    (b) Normalized magnetic flux $\Phi (x,y)$ in the SOL, calculated for a magnetic moment $\vec{\mu}_\mathrm{B}$ at position $\vec{r} = (x,y,0)$ oriented in $z$-direction. The SOL is at $h=\qty{30}{\nm}$ with $\alpha = \qty{10}{\degree}$, corresponding to the tip of the cantilever hovering \qty{5}{\nm} above the sample. Data is normalized to the maximum magnetic flux of \qty{17.4}{\nano \Phi_0}; $\Phi_0$ is the magnetic flux quantum. 
    (c) Out-of-plane component of the Bohr magneton's magnetic field $B_{\mathrm{z}}(x,y)$ at $h=\qty{30}{\nm}$, which is normalized to \qty{68.5}{\nano\tesla}. 
    (d) PSF calculated by iterative Landweber deconvolution from data depicted in (b) and (c) and normalized to its maximum value of \num{0.59}. The outline of the cantilever geometry is shown in dashed lines in (b) and (d).
    }    
    \label{fig:02}       
\end{figure*} 

\subsection{The Coupling Factor}
\label{sec:coupling_factor}

Given the geometry described above, it is possible to simulate a SQUID's response to the stray field that originates from a magnetic particle.
This can be quantified by the coupling factor $\phi_\mu = \Phi/\mu$, which describes how much magnetic flux $\Phi$ per moment $\mu$ is coupled into the SQUID by an idealized point-like magnetic dipole with magnetic moment $\vec{\mu}$, where $\mu = |\vec{\mu}|$ \cite{martinez-perez_nanosquids_2017}.
Naturally, $\phi_\mu(\vec{r},\hat{\mu})$ depends on the location $\vec{r}$ and orientation $\hat{\mu}=\vec{\mu}/\mu$ of the magnetic moment with respect to the SQUID.
Via the superposition principle, the flux coupling into the SQUID from an arbitrary magnetic structure can be calculated by summing up the coupling factors of its constituent magnetic moments and multiplying by the magnitude of each moment: $\Phi = \sum_{i} \mu_i\, \phi_\mu(\vec{r}_i,\hat{\mu}_i)$ \cite{schwarz15_ybco-nanosquids_2015}.
 $\Phi$ depends on the relative position of the SQUID with respect to the magnetic structure, yielding the spatially dependent magnetic flux profile measured via SSM.

Simulations of the coupling factor can be efficiently performed using the Amperian-loop model, which was first described in \cite{Bouchiat2009,Nagel2011} and later confirmed in \cite{Linek2024}.
In this model, the coupling factor for a magnetic moment at position $\vec{r}$ oriented along $\hat{\mu}$ can be calculated as
\begin{align}
    \label{eq:coupling_factor_amperian}
    \phi_\mu(\vec{r}, \hat{\mu}) =  \frac{\vec{B}_J(\vec{r})}{J} \cdot \hat{\mu},
\end{align}
where $\vec{B}_J(\vec{r})$ is the magnetic field produced by the supercurrent $J$, which circulates around the SQUID loop. 
We use the finite element software package 3D-MLSI based on London theory \cite{Khapaev2001,Khapaev2002} to simulate the spatial distribution of the supercurrent density circulating in the SOL.
3D-MLSI calculates discrete sheet current densities in planar 2-dimensional (2D) parallel sheets that are spread over the film thickness $t$. 
To model this film, we use 11 equidistant parallel sheets, which are enough to suppress any spurious results depending on the number of 2D sheets used.
We calculate the 3-dimensional (3D) magnetic field $\vec{B}_J (\vec{r})$ produced by this supercurrent density via the Biot-Savart law.
Results obtained from 3D-MLSI simulations closely match those obtained from Ginzburg-Landau calculations, while reducing simulation time significantly~\cite{Linek2024}.

By using the Amperian-loop model, we avoid calculating the magnetic flux threading through the SQUID loop for every location and orientation of the magnetic moment.
Instead, we simulate the supercurrent distribution around the SQUID loop once. 
This immediately yields the coupling factor at every position and orientation of the magnetic moment via \autoref{eq:coupling_factor_amperian}.

\subsection{The Point Spread Function}
\label{sec:psf}

Using the SOL geometry shown in \autorefsub{fig:02}{a}, we simulate (with $\alpha = \qty{10}{\degree}$) the coupling factor $\phi_\mu(x,y)$ for a magnetic moment pointing in $z$ direction and located in the sample plane with $h=\qty{30}{\nm}$ below the scanning plane. 
In \autorefsub{fig:02}{b}, we then plot the magnetic flux threading through the stationary SQUID $\Phi(x,y)=\mu_\mathrm{B} \phi_\mu(x,y)$ for a magnetic moment carrying one Bohr magneton $\mu_\mathrm{B}$. This expression represents the spatial response of the SOL (at $h=\qty{30}{\nm}$) to a single magnetic moment. 
When compared to the out-of-plane component of the moment's magnetic field $B_z(x,y)$ at $h=\qty{30}{\nm}$, shown in \autorefsub{fig:02}{c}, the broadening and distortion of the flux measured by the SOL compared to the source field is clearly apparent. 
This effect is a consequence of flux focusing effects inherent to the patterned superconducting film that forms the SOL. 
The ability of the SOL to resolve small magnetic features is therefore determined by a combination of the exact geometry of the SOL device and the properties of the superconducting film.

The spatial broadening and distortion of the source magnetic field into the measured flux is described by the point spread function, i.e.\ the convolution of the field with the PSF yields the measured flux: 
\begin{align}
\Phi(x,y) &= B_z(x,y) * \mathrm{PSF}(x,y)\\ \nonumber
&= \iint_{-\infty}^\infty B_z(x',y') \: \mathrm{PSF} (x-x',y-y')\, dx'\, dy'. 
\label{eq:PSF}
\end{align}
The PSF is therefore a property of the probe and is, for non-invasive probes, independent of its distance to the sample or other external factors. 
As both the magnetic flux $\Phi$ and the source field $B_z$ scale linearly with the magnetic moment $\mu$ of the source, the PSF is independent of $\mu$. 
For simplicity, we have chosen $\mu=\mu_\mathrm{B}$; in this case, the flux maps $\Phi(x,y)$ (in units $\Phi_0$) are identical to maps of the coupling factor $\phi_\mu(x,y)$ (in units $\Phi_0/\mu_\mathrm{B}$).
Notably, the PSF carries no units. 
Integrating over its area results in the effective area $A_\mathrm{eff}$ of the SQUID
\footnote{
 If we assume a spatially homogeneous magnetic field $B_z = \mathrm{const.}$, this can be taken out of the integral in Eq.\,(2). Then, the integral over PSF equals $\Phi/B_z$, i.e. the flux induced in the SQUID divided by the applied homogeneous magnetic field, which is the definition of the effective area $A_\mathrm{eff}$ of a SQUID.
 }.

In order to extract the PSF from the simulated magnetic flux in \autorefsub{fig:02}{b} and source stray field in \autorefsub{fig:02}{c}, we must invert the convolution. 
We do so using the Landweber algorithm \cite{Combettes2011}, which was originally developed for deconvolving noisy data with a known filter. 
The Landweber algorithm begins with an initial estimate of the PSF and iteratively reduces the error between the SQUID flux response and the source magnetic field image. 
In Fourier space, the convolution of the source field and the PSF is expressed as a product of the respective Fourier transforms:
\begin{equation}
\mathcal{F}[\Phi](k_x,k_y) = \mathcal{F}[B_z](k_x,k_y) \cdot \mathcal{F}[\mathrm{PSF}](k_x,k_y),
\label{eq:Fourier-PSF}
\end{equation}
where $\mathcal{F}$ denotes the Fourier transform. 
The iterative procedure produces successively better approximations of $(\mathcal{F}[B_z](k_x,k_y))^{-1}$, which can be used to invert \autoref{eq:Fourier-PSF} and approximate $\mathcal{F}[\mathrm{PSF}](k_x,k_y)$. 
The iteration is performed over the index $j \in \mathbb{N}$ using the update rule:
\begin{align}
\begin{split}
\mathcal{F}&[\mathrm{PSF}]^{\,j+1} =\\
&\mathcal{F}[\mathrm{PSF}]^j - \omega \,  \mathcal{F}[\Phi]^* \left( \mathcal{F}[\mathrm{PSF}]^j \cdot  \mathcal{F}[\Phi] - \mathcal{F}[B_z] \right),
\end{split}
\label{eq:Fourier-PSF-k}
\end{align}
where $\omega$ controls the rate of convergence with \linebreak $0 < \omega < 2 \, \max(\mathcal{F}[\Phi])^{-2}$ and $\mathcal{F}[\Phi]^*$ denotes the conjugate transpose of $\mathcal{F}[\Phi]$. 
The iteration proceeds until the residual error becomes comparable to the precision of the simulation or measurement of $\Phi(x,y)$, depending on whether the PSF is extracted from simulation or measurement. 
Finally, an inverse Fourier transform is applied to $\mathcal{F}[\mathrm{PSF}](k_x,k_y)$ to retrieve  $\mathrm{PSF}(x,y)$. \autorefsub{fig:02}{d} shows the resulting PSF, in which the effects of the SOL geometry and flux focusing are clearly visible.

Note that the spectral content of the extracted PSF is limited by the spectral content of the source field $B_z(x,y)$. 
The inversion process cannot recover spectral components of $\mathcal{F}[\mathrm{PSF}]$ if they are not present in $\mathcal{F}[B_z]$. 
Consequently, the smallest reconstructible feature size in the PSF is limited by the smallest features present in the magnetic field $B_z(x, y)$. Therefore, for faithful extraction of a PSF, it is important that the measured source fields are varying over length scales much smaller than the expected PSF.
Moreover, as in any deconvolution, high spatial frequency components are highly sensitive to noise. These components are gradually amplified in the Landweber algorithm with increasing iteration count, leading to amplification of the noise and introduction of artifacts. 
Therefore, the iteration must be terminated early enough to balance convergence to the true PSF and the appearance of high-frequency artifacts.

\begin{figure*} 
    \includegraphics[width=\textwidth]{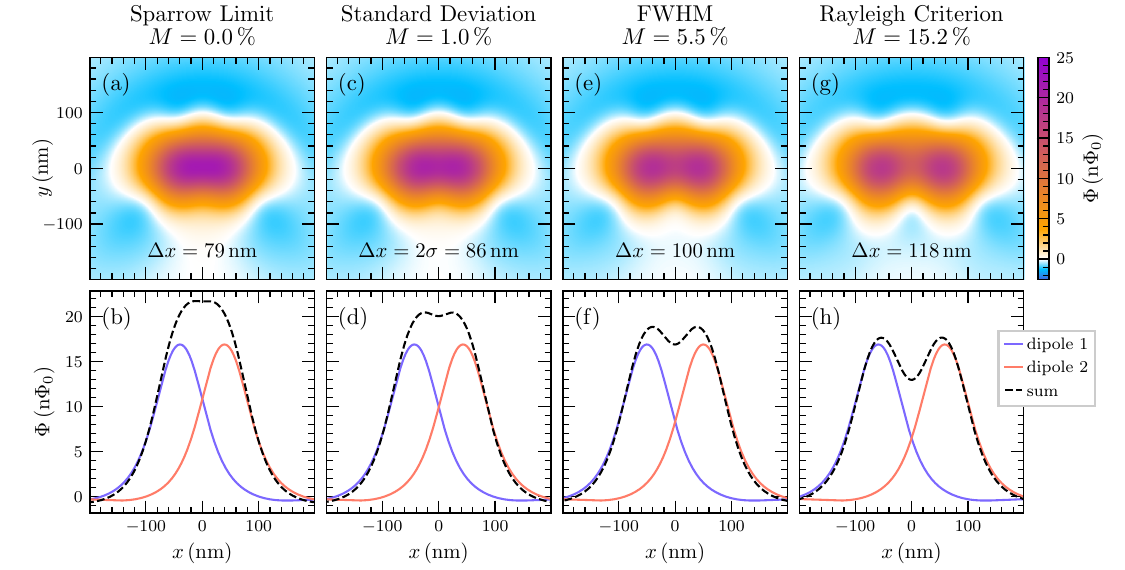}
    \caption{Simulated magnetic flux $\Phi(x,y)$ (upper row) and corresponding line scans $\Phi(x,y=0)$ (bottom row) coupled to the same SOL geometry as in \autoref{fig:02} (for $\alpha = 0$) at height $h=\qty{30}{\nm}$ above two dipoles with moments $\mu=\mu_\mathrm{B}$ along the $z$-direction. The moments are placed at different distances $\Delta x$ to illustrate several spatial resolution criteria.
    The line scans show the signals from the individual dipoles (solid lines) and their sum (dashed line).
    The standard deviation $\sigma = \qty{43}{\nm}$ is acquired by fitting a Gaussian to the individual dipole signal. 
    }
    \label{fig:03}
\end{figure*} 

\begin{figure*} 
    \includegraphics[width=\textwidth]{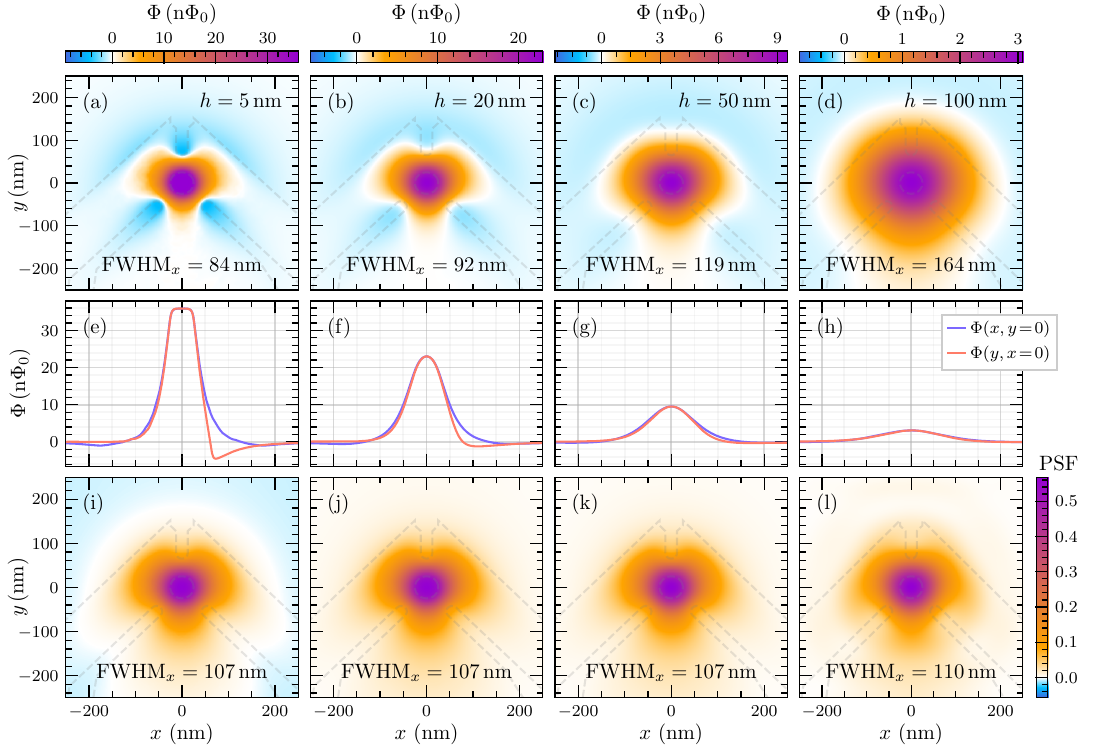}
    \caption{Simulated magnetic flux $\Phi(x,y)$ (upper row), corresponding line scans $\Phi(x,y=0)$ and $\Phi(x=0,y)$ (middle row) and extracted PSF (bottom row) for the same SOL geometry as in \autoref{fig:02} (for $\alpha=0$) at different heights $h$ increasing from \qty{5}{\nm} (left column) to \qty{100}{\nm} (right column) above a single dipole with moment $\mu=\mu_\mathrm{B}$ along the $z$-direction.
    }
    \label{fig:04}
\end{figure*} 

\begin{figure} 
    \centering
    \includegraphics[width=\columnwidth]{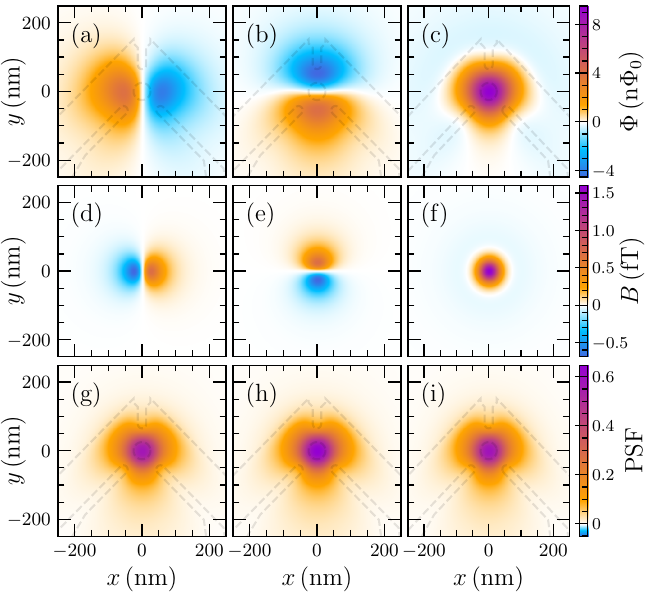}
    \caption{Calculated characteristics for the same SOL geometry as in \autoref{fig:02} (for $\alpha=0$), at height $h=\qty{50}{\nm}$, induced by magnetic dipoles with moment $\mu_B$ oriented along $x$ (left), $y$ (center) and $z$ (right) directions.
    (a)-(c) Magnetic flux in the SQUID. (d)-(f) $z$-component of the magnetic dipole field at $h=\qty{50}{\nm}$. (g)-(i) extracted PSFs.
    }
    \label{fig:05}
\end{figure} 

\begin{figure}[t!] 
    \includegraphics{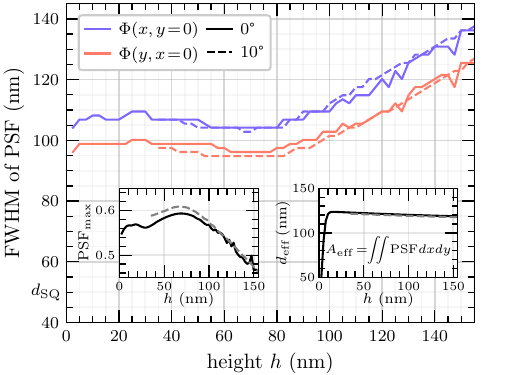}
    \caption{FWHM of the PSF$(x,y=0)$ and PSF$(x=0,y)$ vs. height $h$ above a magnetic dipole oriented in $z$ direction, calculated for the same SOL geometry as in \autoref{fig:02} at $\alpha=\qty{0}{\degree}$ (solid lines) and $\qty{10}{\degree}$ (dashed lines). 
    Insets show the maximum height of the PSF vs. $h$ (left) and the effective diameter, calculated by integrating over the PSF which results in the effective area (right).
    }
    \label{fig:06}
\end{figure} 

\section{Spatial Resolution}
\label{sec:sol_dipol_simulations}

The spatial resolution of a probe is commonly defined as the smallest separation at which distinct features can be reliably distinguished. 
Simulations of the response to point-like magnetic dipoles are thus ideally suited for probing the spatial resolution of images taken by a given SOL geometry. 
\Autoref{fig:03} shows simulations of $\Phi(x,y)$ produced by two nearby magnetic moments in an SOL with diameter $d_\mathrm{SQ} = \qty{50}{\nm}$ at a height $h=\qty{30}{\nm}$. 
Spacings $\Delta x$ between the two moments that correspond to different spatial resolution criteria are shown, including the Sparrow limit and Rayleigh criterion. 
The Sparrow limit is defined as the separation below which the minimum in the response curve from two point-like features disappears, i.e.\ when the modulation depth $M=0$. 
The Rayleigh criterion corresponds to the separation below which $M$ reduces to less than $\qty{15.2}{\percent}$ \cite{valenzuela2019basic, sirohi2021resolving}. 
Criteria based on the width of the response to a point-like source, i.e.\ when $\Delta x$ equals the full width at half maximum (FWHM) or twice the standard deviation $2\sigma$, are also shown. 

Having been developed for optics, these criteria are ill-suited to SSM and other scanning probe microscopies where the PSF can have a complex structure and may not be approximated by a Gaussian.
Although often valid in the far field, in the near field, where the distance between the probe and the source is comparable to the size of the PSF (as is apparent in \autorefsub{fig:02}{d} or \autorefsub{fig:04}{i} to (l)) such measures begin to fail. 
In this limit, criteria may depend on the direction along which they are evaluated and do not capture the image distortion produced by the probe.

Given that these criteria provide characteristic lengths of the probe's response to a point-like magnetic source, they also depend on the probe-sample distance. 
In the appendix, \autoref{fig:08} shows how the FWHM criterion changes as a function of $h$ for SOLs of various diameters $d_\mathrm{SQ}$. 
The spatial resolution is seen to improve with reduced probe-sample distance, only saturating at a limit set by the characteristic size of the SOL, $d_\mathrm{SQ}$.
Furthermore, we can see that decreasing the SQUID hole diameter reduces the width at which the signal saturates, therefore improving spatial resolution.

A more rigorous and quantitative method for describing a probe's ability to resolve spatial features is provided by the PSF, which in the case of the SOL provides a flux map of the probe's response to a magnetic field source. 
Unlike the aforementioned criteria, the PSF is purely a property of the probe and is independent of the source magnetic field or the distance of the probe from the source (as long as it is a non-invasive probe). 
Therefore, its characteristic lateral extent, e.g.\ its FWHM, is a measure of the intrinsic spatial resolution of a given probe.
Because the SQUID lacks rotational symmetry due to the presence of the Josephson junctions, the PSF, and hence its FWHM, is anisotropic. 
Therefore, the spatial resolution of the SQUID varies for different scanning directions.

To demonstrate the suitability of the FWHM of the PSF as a spatial resolution criterion, \Autoref{fig:04} shows PSFs (bottom row) extracted by deconvolving the simulated magnetic flux maps (upper row) with the source magnetic field produced by a magnetic moment $\mu_\mathrm{B}$ along the $z$-direction for several sensor-sample separations $h$. 
The magnetic flux line scans (middle row) illustrate the distance dependence of the flux profile amplitude and FWHM. In contrast, the FWHM of the extracted PSF remains invariant at $\approx \qty{107}{\nm}$ over the investigated height range from $h \in (\qty{5}{\nano\meter}; \qty{100}{\nano\meter})$, confirming that it constitutes an intrinsic, geometry-limited resolution metric. 

In \autoref{fig:05} we additionally show PSFs calculated from three different orientations of a dipole along $x$, $y$ and $z$-direction resulting in completely different stray fields and magnetic flux coupling into the SQUID.
Nevertheless, the PSFs remain identical for all three orientations of the dipole, which demonstrates the independence of the PSF from applied magnetic fields and proves the effectiveness of our method.
This again emphasizes that the PSF is a sensor-intrinsic value which is not influenced by the change of magnetic fields.  

To further strengthen the argument, we show the FWHM of PSF line scans as a function of height $h$ in \autoref{fig:06}.
The PSF width is almost constant for $h \lesssim \qty{80}{\nano\meter}$ in both the $x$ and $y$ directions. 
For $h \gtrsim \qty{80}{\nano\meter}$, the PSF width increases almost linearly with increasing $h$, because of a failure of the reconstruction algorithm: beyond this probe-sample distance, the magnetic field profile of the source has spread out to the extent that its smallest spatial features begin to exceed the size of the probe's spatial features. 
As a result, the deconvolution algorithm cannot recover the true PSF and its size becomes limited by the spreading of the source field, as discussed in \autoref{sec:psf}.

The saturation of the FWHM of the PSF for $h \lesssim \qty{80}{\nano\meter}$ also indicates that the signals from a single magnetic dipole have enough frequency information to fully describe the width of the PSF.
By comparing calculations for tilt angles $\alpha=\qty{0}{\degree}$ (solid lines) and $\qty{10}{\degree}$ (dashed lines) in \autoref{fig:06}, we find that moderately tilting the cantilever has little to no effect on the width of the PSF.

As mentioned in Sec.\ref{sec:psf}, the integral over the PSF yields $A_\mathrm{eff}$ of the SQUID, which can be converted into an effective diameter $d_\mathrm{eff}=2\sqrt{A_\mathrm{eff}/\pi}$. 
For the SQUID geometry shown in \autoref{fig:02}, we determine $A_\mathrm{eff}$ with 3D-MLSI by calculating the current density for a constant applied magnetic field to be $d_\mathrm{eff} = \qty{118}{\nm}$. 
As a consistency check, the right inset in \autoref{fig:06} shows $d_\mathrm{eff}$, determined from the integral over the PSF, vs. height. 
In excellent agreement, we observe an almost constant $d_\mathrm{eff}\approx\qty{120}{\nm}$ over a wide range of $h$ down to \qty{10}{\nm}. 
Even for $h \gtrsim \qty{80}{\nano\meter}$, $d_\mathrm{eff}$ remains roughly constant, as the increase in FWHM with $h$ is compensated by the decrease in the maximum value PSF$_\mathrm{max}$ of PSF$(x,y)$ (see left inset in \autoref{fig:06}). 
We ascribe the strong decrease of $d_\mathrm{eff}$ for very small $h\lesssim \qty{10}{\nano\meter}$ to artifacts which arise from slightly negative values for the PSF far outside the SQUID loop (c.f.~\autorefsub{fig:04}{i}).

We also investigate the influence of the London penetration depth $\lambda_\mathrm{L}$ on the SOL's spatial resolution.
As shown in the appendix, in \autoref{fig:09}, varying $\lambda_\mathrm{L}$ on the order of several tens of nanometers produces no measurable change in the FWHM of the sensor's signal response. 
This independence arises because the critical structural dimensions of the thin film SQUID geometry, most notably the junction widths and film thickness, are smaller than $\lambda_\mathrm{L}$. In this regime, the flux coupling is fundamentally dictated by the physical geometry of the device.

\section{Imaging a skyrmion}
\label{sec:skyrmions}

\begin{figure*}[t!] 
    \includegraphics[width=\textwidth]{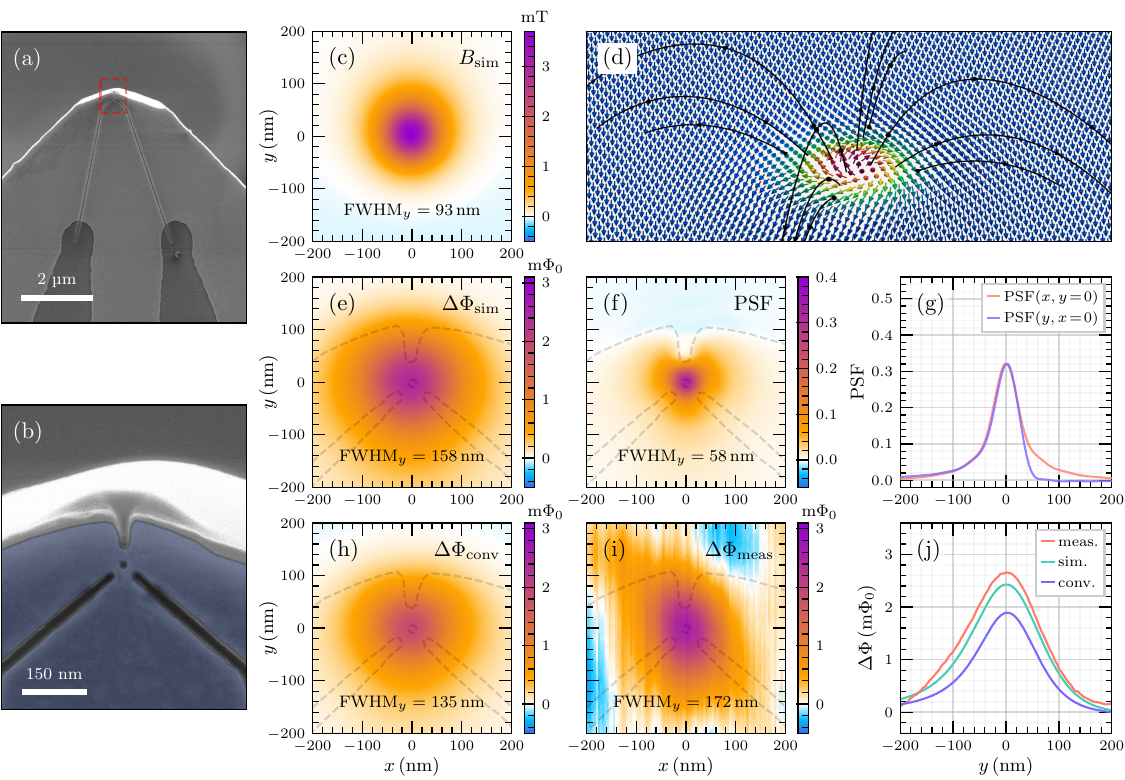}
    \caption{Comparison of calculated and measured magnetic flux signals of the SOL induced by a skyrmion. 
    (a), SEM image (top view) of a \ce{Nb} SQUID on a planar Si cantilever used for the imaging study. The dashed red frame indicates the zoom area shown in (b), where blue regions indicate the extracted shape of the \ce{Nb} film, used for simulations of $d_\mathrm{eff,sim}$ and $\phi_\mu$. 
    (c) Simulated magnetic stray field of a skyrmion in a $\qty{200}{\nm}$ thick bulk \CuOSeO sample at \qty{10}{\degree} tilt angle \qty{50}{\nm} above the sample. 
    (d) Corresponding spin configuration (saturation magnetization) in the top layer of the skyrmion phase in the bulk.
    (e) Simulated magnetic flux change $\Delta\Phi_\mathrm{sim}$ at a height $h=\qty{50}{\nm}$ and angle $\alpha = \qty{10}{\degree}$ calculated from the skyrmion's magnetization configuration and the simulated coupling factor $\phi_\mu$ of the SOL. 
    (f) PSF calculated from magnetic field of a single dipole and the simulated SQUID response to a single dipole, using Landweber iteration. 
    (g) Line scans of the PSF along $x$ direction (at $y=0$, red line) and along $y$ diretion (at $x=0$, blue line).
    (h) Simulation of the magnetic flux $\Delta\Phi_\mathrm{conv}$ at a height $h=\qty{50}{\nm}$ and angle $\alpha = \qty{10}{\degree}$, by convoluting the stray field of a skyrmion (c) with the PSF calculated in (f).  
    (i) Magnetic flux change $\Delta\Phi_\mathrm{meas}$ in the experimental scan of the skyrmion phase on the surface of a bulk \CuOSeO sample at $h=\qty{50\pm10}{\nm}$ and $\alpha = \qty{10}{\degree}$.
    (j) Line scans along $y$-direction (at $x=0$) of magnetic flux changes from (e), (h) and (i), i.e., for both simulation methods and the measured data.
    For comparison, the projection of the cantilever geometry onto the $x$-$y$ plane is illustrated in grey in (e,f,h,i).
    }
    \label{fig:07}
\end{figure*} 


In order to verify our method of calculating the signal response of our SOLs and therefore the PSFs even further, we carry out an experiment with a real SOL and image a magnetic field profile with small characteristic sizes. 
Using one of the smallest Nb SOLs fabricated thus far, pictured in scanning electron microscopy (SEM) images \autorefsub{fig:07}{a} and (b), we map the flux contrast $\Delta \Phi (x,y)$ produced by an isolated magnetic skyrmion at the surface of a bulk \CuOSeO crystal, as also reported by Weber et al.~\cite{Weber2025}. 
At low temperature, this insulating cubic helimagnet hosts a number of modulated magnetic phases, including low-temperature skyrmion (LTS) and helical (H) phases, which produce nanometer-scale magnetic field patterns at the sample surface~\cite{Marchiori2024}. 
In particular, in magnetic fields applied along $\langle 100 \rangle$, previous SSM has shown that the LTS phase appears on the corresponding $\{100\}$ surface in the form of clusters of disordered skyrmions within a field-polarized (FP) background. 
In images of the magnetic field, individual skyrmions generate a point-like reduction in the out-of-plane stray-field, as a result of their core magnetization opposing the surrounding FP phase. 
According to micromagnetic simulations, skyrmions in \CuOSeO have a radius of $\sim\qty{10}{\nm}$, corresponding to the distance from their center, where their magnetization opposes the FP background, to where their out-of-plane magnetization vanishes.

In order to experimentally image an isolated skyrmion in the LTS phase, we initially saturate the system with an out-of-plane applied magnetic field along $[001]$ of $\mu_0 H_z = \qty{200}{\milli\tesla}$ at $T = \qty{5}{\kelvin}$. 
We then reverse this field, and ramp to \qty{-160}{\milli\tesla}, as described in Weber et al.~\cite{Weber2025}. 
At this point, the surface of the \CuOSeO is populated by clusters of skyrmions within an FP background. 
Using the SOL, we then map $\Delta\Phi_\mathrm{meas} (x,y)$, produced by the stray field of a single skyrmion at the $(001)$ surface threading through the SQUID loop. \autorefsub{fig:07}{i} shows the flux map of a single skyrmion, measured at a constant tip-sample spacing of $50 \pm$\qty{10}{\nm}. 
We show here only the change in magnetic flux induced by the skyrmion, since the SQUID measures only relative changes in magnetic flux.

We next simulate the stray field produced by such an isolated skyrmion. 
We model its magnetization and the resulting $\vec{B}_\mathrm{sim}(x,y)$ that it produces, using the micromagnetic software package \textit{MuMax3}, which is based on the Landau-Liftshitz-Gilbert formalism~\cite{Vansteenkiste2014,Exl2014}. 
As simulation parameters, we use a saturation magnetization of \qty{103}{\kilo\ampere/\meter}, an exchange stiffness of \qty{0.42}{\pico\joule/\meter}, a bulk Dzyaloshinskii-Moriya constant of \qty{0.89}{\micro\joule\per\meter^2}, and a cubic anisotropy constant of \qty{1.2}{\kilo\joule\per\meter^3}.
The corresponding stray field of a single skyrmion at the probe-sample distance of $h=\qty{50}{\nm}$ is presented in \autorefsub{fig:07}{c} together with a schematic representation of the magnetization in the top layer in \autorefsub{fig:07}{d}. 

We then model the shape of the SQUID shown in \autorefsub{fig:07}{b}. A thickness of $t=\qty{50}{\nm}$ with a total of 11 superconducting simulation layers is used. 
We use a SQUID hole diameter $d_\mathrm{SQ} = \qty{15}{\nano\meter}$, which follows the size indicated in the SEM image, as we assume low edge damage during the \ce{He}-FIB-milling process.
With the digitized geometry we simulate an effective diameter of $d_\mathrm{eff,sim}=\qty{64}{\nm}$ similar to the measured effective diameter $ d_\mathrm{eff,meas}\approx \qty{62}{\nm}$ \cite{Weber2025}.
Moreover, we simulate coupling factors $\phi_\mu$ with \autoref{eq:coupling_factor_amperian}.
Together with the known magnetization configuration of the skyrmion, it is possible to simulate the response of the SOL to the skyrmion.
For this purpose, we assume a magnetic dipole density corresponding to a saturation magnetization of \qty{103}{\kilo\ampere/\meter} and compute the magnetization within a simulation volume of \qty{3}{\micro\meter} × \qty{3}{\micro\meter} × \qty{200}{\nano\meter} (width, length, and height, respectively).
By discretizing the magnetization into voxels, we then use the coupling factor $\phi_\mu(x,y,z)$ at the position of each magnetic moment together with the value and diretion of the magnetic moment to calculate the full magnetic flux response $\Phi_\mathrm{sim}(x,y)$ to the skyrmion's simulated magnetization pattern, shown in \autorefsub{fig:07}{e}. 

As in previous discussions in \autoref{sec:psf} we calculate the PSF of the SQUID geometry in \autorefsub{fig:07}{f} from the simulated magnetic field of and signal response to a single dipole at $h=\qty{50}{\nm}$ using the Landweber algorithm.
We can again verify this PSF by summing over it and retrieving an effective diameter of \qty{66}{\nm}, close to the simulated and measured effective diameters.
\autorefsub{fig:07}{g} shows line scans through the maximum of the PSF along the $x$- and $y$-directions.
We demonstrate that the FWHM of the PSF along the $x$-direction is \qty{63}{\nm} and along the $y$-direction \qty{58}{\nm}, thereby verifying previous claims about the spatial resolution of our probes \cite{Weber2025}.
As shown previously, the anisotropy in the PSF hints towards better spatial resolution along the $y$-direction due to the implementation of a third junction by adding a cut from the tip of the cantilever.

We do not directly extract the PSF from the simulated magnetic field of the skyrmion (\autorefsub{fig:07}{c}) and the simulated or measured flux response of the SQUID, because the skyrmion field does not contain high enough spatial frequencies to accurately reconstruct the PSF.
Nevertheless, the PSF calculated from the single-dipole field is validated by convoluting it with the skyrmion field in \autorefsub{fig:07}{c}, resulting in yet another SOL magnetic flux response $\Delta\Phi_\mathrm{conv}(x,y)$ to the skyrmion, shown in \autorefsub{fig:07}{h}.
Note that this signal is weaker than the directly calculated flux image of the skyrmion in \autorefsub{fig:07}{e}.
We attribute this decrease in signal strength to a lack of low-frequency components in the PSF, which cannot be extracted from the dipole simulations. 
The extraction of spatial resolution from the PSF is only slightly affected by this effect, because their spectral weight in the total response is small compared to that of higher-frequency components.

To summarize, we first simulate the SOL response $\Delta \Phi_\text{sim}$ to the skyrmion by calculating coupling factors and thereby adding up the contribution to the total flux of each magnetic moment making up the skyrmion. 
Second, we calculate the SOL response $\Delta \Phi_\text{conv}$ to the skyrmion using a PSF determined from simulations of its response to a single magnetic dipole and from the simulated stray field of the skyrmion. 

Finally, in \autorefsub{fig:07}{j} we compare these two methods with the measured magnetic flux contrast of a skyrmion $\Delta \Phi_\text{meas}$.
We see that the SQUID signal simulated by the coupling factors matches the measured data of the skyrmion.
The measured FWHM of the skyrmion signal is slightly larger than simulated, probably as a result of the lack of an absolute reference in the measured data.
The flux calculated via the PSF has a weaker signal strength and smaller FWHM than the other two flux profiles, reflecting the missing low-frequency components in the PSF, which were lost in its extraction. Nevertheless, the FWHM of the PSF can still be verified to be far below \qty{70}{\nm},
demonstrating the great potential of this method to calculate the PSF for an SOL geometry.

Several further sources of error exist that can explain the discrepancies between measurements and simulation.
First, the geometry of the SQUID can never be perfectly reconstructed due to unknown effects of edge damage. 
Additionally, the SQUID's \ce{Nb} film thickness is only known to a certain degree, again affecting the coupling to the SQUID.
Other sources of error include the cantilever's scanning height, which is only known by $\pm\qty{10}{\nm}$, and uncertainties in the properties of the \CuOSeO sample.
Moreover, the noise background and the signal of surrounding skyrmions in the measurements of the skyrmions make it hard to align the background signal of simulations and measurements.

\section{Conclusion}
\label{sec:conclusion}
In this work, we model the spatial response of nanometer-scale probes for SSM. For a given SOL geometry, we determine the PSF, which provides a full description of the probe sensitivity, and we introduce the width of the PSF as an appropriate intrinsic measure of spatial resolution in SSM.
Moreover, the PSF can be used to deconvolve the measured spatial response of the probe to a source into an image of the source, a crucial step for interpreting an image.

By simulating the magnetic coupling between a point-like magnetic dipole and the SOL, we extract the PSF and demonstrate that the calculated spatial response of the magnetic flux signal accurately reproduces an experimental result, obtained from imaging an individual magnetic skyrmion in bulk \CuOSeOdot. Ultimately, we verify that our SOL probes achieve spatial resolutions well below \qty{100}{\nm}, exhibiting a full width at half maximum of the PSF below \qty{70}{\nm} in the $x$- and below \qty{60}{\nm} in the $y$ directions.

While these results firmly establish the high-resolution capabilities of SOL sensors, significant potential remains for future optimization. Currently, spatial resolution is primarily limited by flux focusing effects, that cause a broadening of the shape of the PSF. 
However, the mathematical framework presented here provides a direct pathway to computationally optimize the geometry of the SOL. 
Beyond simply reducing the SQUID hole size, we could systematically investigate the shape of the FIB cuts, adjust the superconducting film thickness, and vary cantilever tilt angle to further enhance spatial resolution. 
We expect the most significant improvement by removing superconducting material, to further narrow down the width of the SQUID loop geometry. However, this will inevitably increase the SQUID inductance and hence the intrinsic thermal white noise of the SQUID.
Accordingly, integrating noise models for the SOL is a crucial next step, as the signal-to-noise ratio fundamentally impacts effective resolving power. 

\begin{acknowledgments}
We thank P. R. Baral and A. Magrez for growing the \CuOSeO sample and Ute Drechsler, Armin Knoll and Olvier Kieler for their contributions to the SOL fabrication. We acknowledge support by the European Commission under H2020 FET Open grant “FIBsuperProbes” (Grant No. 892427) and the Swiss National Science Foundation under Grant No. 200020-207933. 
We also gratefully acknowledge support by the COST actions NANOCOHYBRI (CA16218), FIT4NANO (CA19140) and SUPERQUMAP (CA21144).
\end{acknowledgments}

\section*{Data Availability}
The data that support the findings of this article are not publicly available. The data are available from the authors on reasonable request.

\appendix

\section{Width of magnetic flux distribution (coupling factor) vs.~SQUID hole diameter}

\begin{figure} 
    \includegraphics{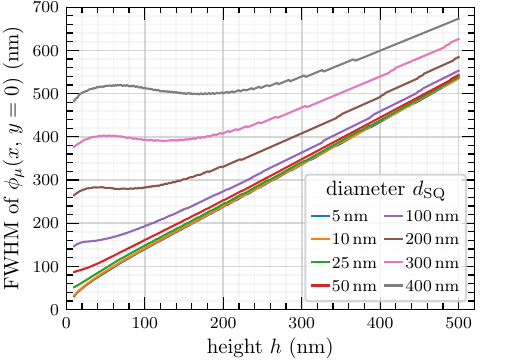}
    \caption{FWHM of the coupling factor $\phi_\mu(x,y=0)$ of the SOL as in \autoref{fig:02} ($\alpha = 0$, $\lambda_\mathrm{L}=\qty{108}{\nm}$) with a magnetic dipole along $z$-direction vs.~scanning height $h$ for different SQUID hole diameters $d_\mathrm{SQ}$.
    }
    \label{fig:08}
\end{figure}

We perform simulations of the FWHM of the coupling factor $\phi_\mu(x, y=0)$ vs.~scanning height $h$ for different SQUID hole diameters $d_\mathrm{SQ}$ from \num{5} to \qty{400}{\nm}, which are shown in \autoref{fig:08}.
The same SOL layout as in \autoref{fig:02} was used, coupled to a single dipole along the $z$-direction.
To reduce simulation time, this data is only simulated with two layers in the superconducting layer of the SOL, which has little effect on the FWHM of the coupling factor.
For larger SQUID diameters, the FWHM of the signal response saturates (with decreasing $h$) at a value close to $d_\mathrm{SQ}$. 

\section{Width of magnetic flux distribution (coupling factor) vs.~London penetration depth}

\begin{figure}[t!]
    \includegraphics{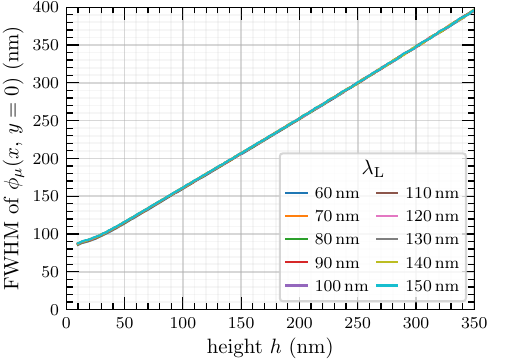}
    \caption{ FWHM of the coupling factor $\phi_\mu(x,y=0)$ of the SOL as in \autoref{fig:02} ($\alpha = \qty{0}{\degree}$, $d_\mathrm{SQ}=\qty{50}{\nm}$) with a magnetic dipole along $z$-direction vs.~scanning height $h$ for different London penetration depths $\lambda_\mathrm{L}$. All lines are displayed above each other. }
    \label{fig:09}
\end{figure}

As the relevant feature sizes in the thin film SQUID geometry are smaller than the London penetration depth $\lambda_\mathrm{L}$, it does not significantly affect flux coupling.
In the limit of $\lambda_\mathrm{L}$ being much larger than lengths over which the geometry of the superconducting device changes, the supercurrent distribution in the device becomes homogeneous and flux coupling does not depend on $\lambda_\mathrm{L}$.
In \autoref{fig:09} the FWHM of the coupling factor $\phi_\mu(x,y=0)$ vs.~scanning height $h$ is shown for values of $\lambda_\mathrm{L}$ from \num{60} to \qty{150}{\nm}. 
The same SOL layout as in \autoref{fig:02} is used, with $d_\mathrm{SQ}=\qty{50}{\nm}$. 
As above, these simulations have been performed with only two superconducting layers which has an effect on the maximum signal of the SQUID but not on the width. 
All curves in \autoref{fig:09} fall on top of each other, i.e., the width of the flux signal are unaffected by $\lambda_\mathrm{L}$.
\vspace{1.3 cm}

\newpage

\bibliography{literature}

\end{document}